\documentstyle[aasms4]{article}

\def\arcsec{\hbox{$^{\prime\prime}$}}
\def\kms{km s$^{-1}$}

\begin{document}

\pretolerance=100
\tolerance=500
\rightskip=0pt

\title{FCC 35 and its \ion{H}{1} Companion:  Multi-Wavelength Observations and
Interpretation} \author{ M.E. Putman\altaffilmark{1}, M.
Bureau\altaffilmark{1}, J.R. Mould\altaffilmark{1}, L.
Staveley-Smith\altaffilmark{2}, and K.C. Freeman\altaffilmark{1}}
\altaffilmark{1}{Mount Stromlo and Siding Spring Observatories,
Institute of Advanced Studies, The Australian National University,
Private Bag, Weston Creek Post Office, ACT 2611, Australia}

\altaffilmark{2}{Australia Telescope National Facility, CSIRO, P.O. Box
76, Epping, NSW 2121, Australia}

\begin{abstract} \rightskip=\leftskip
The Fornax cluster galaxy FCC 35 shows an unusual multiply-peaked
integrated \ion{H}{1} profile (Bureau, Mould \& Staveley-Smith 1996).  We
have now observed FCC 35 with the Australia Telescope Compact Array
(ATCA) and have found a compact \ion{H}{1} source with $M_{HI}$ = 2.2 $\times$
$10^{8}$ $M_{\odot}$, and a spatially overlapping complex of \ion{H}{1} gas with the same mass. 
  By combining optical observations with
the \ion{H}{1} data, we are able to identify FCC 35 as a young compact source of star
formation with a nearby intergalactic \ion{H}{1} cloud which is devoid of
stars.  We classify FCC 35 as a blue compact dwarf (BCD) or \ion{H}{2}
galaxy, having large amounts of neutral hydrogen, very blue colors
($(U-V)$ = 0.1), and a low metallicity spectrum with strong narrow
emission lines.   Together with the presence of the \ion{H}{1} cloud, this
suggests that FCC 35 is the result of a recent interaction within the
Fornax cluster. 
\end{abstract} 

\clearpage
\section{Introduction}

Our interest in FCC 35 began in 1994 when this galaxy was observed as
part of an investigation of the Tully-Fisher relation in Fornax (Bureau, 
Mould, \& Staveley-Smith 1996; hereafter BMS).  Optically, FCC 35 was 
identified as a member of the Fornax cluster by Ferguson (1989) and 
classified as a possible BCD/Sm IV.  BMS photometry revealed a high surface 
brightness and an offset nucleus, despite a regular light profile.  The 
21 cm single-dish observations of FCC 35 showed three distinct \ion{H}{1} peaks
within a 700 km s$^{-1}$ velocity range (see Fig.~\ref{fig:f1}).   There 
were no other known \ion{H}{1} sources within the Parkes' beam, so the explanation 
for the three-peaked profile was unknown.  This anomaly served to motivate 
further studies of FCC 35.

Generally, BCDs like FCC 35 are high surface brightness dwarf galaxies
which appear to be undergoing an intense period of star formation
(Thuan \& Martin 1981).  The cause of these star formation bursts is not well understood,
but in many cases interaction is considered a likely mechanism.  Interactions can
induce rapid star formation (Bushouse 1987) and the resulting internal
motions can continue to induce bursts for up to 10$^8$ yrs after the 
interaction (Noguchi 1991).  There are several types of interaction
which can trigger the BCD phenomenon.  The first involves an encounter
between two spiral galaxies and the formation of \ion{H}{1}-rich tidal tails. 
 BCDs have been observed and modeled to form at the end of these tails which can extend
for hundreds of kpc (e.g. Duc et al. 1997; Barnes \& Hernquist 1992, 1996; Elmegreen, Kaufman \& Thomasson
1993; Mirabel, Lutz \& Maza 1991; Schweizer 1978).   The surrounding environment is often left in a
disordered state for some time after this type of interaction.  Another
type of starburst-inducing interaction involves an intergalactic \ion{H}{1}
cloud of similar mass to the progenitor galaxy (Taylor, Brinks \&
Skillman 1993, Taylor 1997).  Taylor et al. (1995, 1996) completed a survey of relatively
isolated BCDs (also called H~{\footnotesize II} galaxies) and found $\approx$57\% of
these to have an \ion{H}{1} companion which could be triggering the
 star formation burst.  Finally, close encounters between two galaxies of
different mass often induce star formation in the smaller component,
possibly creating a BCD (Ostlin \& Bergvall 1993; Lacey \& Silk
1991).  Putting all of these possibilities together strongly suggests
that the star formation bursts which produce BCDs are indeed due to
interactions.

In a cluster environment these interactions are much more
likely to occur than in the field.  The denser the
environment, the higher the potential for an interaction among member
galaxies.  Fornax is a well-studied nearby cluster (d = 18.2 Mpc;
Madore et al. 1996) which has one of the highest galaxy volume
densities in the Local Supercluster (Held \& Mould 1994) and more than
twice the central surface density of Virgo (Ferguson \& Sandage
1988).  The relatively low velocity dispersion of Fornax
($\approx$ 400 \kms) further favors interaction between cluster
members. FCC 35 is therefore in an ideal environment to be affected by the mechanisms
described above.

The ATCA observations of FCC 35, presented here, reveal two distinct \ion{H}{1}
sources; one compact and regular associated with the optical FCC 35, and one
extended and irregular with no optical counterpart.  The sources
overlap spatially but are separated by 140 km s$^{-1}$ in velocity.
These observations indicate that FCC 35 has an \ion{H}{1} companion of comparable
mass.  This brings us to the various interaction scenarios.  The
starburst which is now FCC 35 may be due to a previous interaction with
this \ion{H}{1} source or both components could be the result of a
spiral-spiral interaction.  It is also conceivable that the \ion{H}{1}
companion itself resulted from a gas outflow associated with the
star formation burst. Investigating these possibilities is one of the central
topics of this paper.

In this paper we present a combination of data which helps to
reveal the origin and evolution of FCC 35.  We discuss the ATCA
observations and data reduction in $\S$2.1 and the optical
imaging and spectroscopy in $\S$2.2.  In $\S$3.0 we present the
results of these observations, including the \ion{H}{1} distribution ($\S$3.1),
\ion{H}{1} kinematics ($\S$3.2), stellar distribution ($\S$3.3), and physical
conditions of the gas ($\S$3.4).  Finally, in $\S$4.0, we discuss these results
and their implications for the formation and evolution of dwarf galaxies, in
particular with respect to various interaction scenarios.  The 21 cm
and optical observations together provide a unique source of
information on the nature of BCDs and their
companions in clusters.  

\section{Observations}
\subsection{Radio Synthesis Observations}

21 cm line data were obtained with the ATCA using three configurations and
all six antennas.  The observing log can be found in Table~\ref{tab:t1}.   
The total bandwidth of the spectral line data was 8 MHz or 1688 km s$^{-1}$, encompassing all
three of the peaks which were detected in the Parkes observations
(see Fig.~\ref{fig:f1}).  The central velocity was 1550 km s$^{-1}$,
with 512 channels in each polarization spaced by 3.3 km s$^{-1}$.   Continuum 
data centered at 1380 MHz was also obtained during the ATCA observations, but FCC 35 
remained undetected at the 0.45 mJy level.  

The spectral line
data were loaded and immediately Hanning smoothed in MIRIAD, resulting
in 256 channels at a velocity resolution of 6.6 km s$^{-1}$.  Bad data
were flagged interactively by examining both the phase and amplitude
for suspect times and baselines.  The bandpass and flux level were
calibrated using the primary calibrator PKS B1934-638, which was
observed at the beginning of each run.  The gain and phase were
calibrated using the secondary calibrators listed in Table~\ref{tab:t1}; 
these were observed every 45 minutes during the observations (flux reference:
Walsh 1992).

After calibration, the continuum was subtracted in the {\em uv} plane
using a second order fit to the line free channels.
 The phase center of the visibility data was moved to the
strong continuum source Fornax A in order to achieve a more accurate continuum subtraction.  The {\em uv} data
for each configuration were then combined and imaged using natural weighting, a cell-size of
15\arcsec $\times$ 15\arcsec, and {\em uv} spacings out to 8
k${\lambda}$.  The final cleaned cube has 128 $\times$ 128 pixels and
 140 channels, with a resulting beam FWHM of 60\arcsec $\times$
40\arcsec.  The rms noise per channel is 1.2 mJy/beam. 
A uniformly weighted cube of the same size was also created
using a cellsize of 5\arcsec $\times$ 5\arcsec\  and a {\em uv} taper of 9
k$\lambda$.  The beam FWHM in this case is 30\arcsec $\times$
25\arcsec\  and the rms noise per channel is 2 mJy/beam.

\subsection{Optical Observations}

V and I band images were obtained at the f/8 focus of the 1.0 m telescope at
Siding Spring Observatory (SSO).  A U band image was subsequently obtained
using the Imager on the SSO 2.3 m telescope.  A log of the optical observations can be
found in Table \ref{tab:t2}.  The scale of the images is
~0\farcs6 pixel$^{-1}$ and all observations were done under
photometric conditions.  The data reduction was completed within the IRAF
package using standard procedures for bias subtraction and flatfielding
with sky flats.

Long-slit optical spectra were also obtained using the Double Beam
Spectrograph (DBS) also on the SSO 2.3 m telescope. The slit was
positioned along the major axis of FCC 35 and a slit width of
1\farcs8 was used, yielding a two-pixel spectral resolution of
1.1~\AA. The pixel scale in the spatial direction is 0\farcs9
pixel$^{-1}$. The data were reduced in the standard manner within IRAF.
The data were first bias subtracted and flatfielded, and then
wavelength calibrated using bracketing comparison lamp exposures. The data
were then rebinned to a linear wavelength scale, slit corrected, and
sky and continuum subtracted to ``isolate'' emission lines. Because the
resulting spectra did not show any obvious velocity gradient or
variation in line ratios across the galaxy, the data were collapsed
spatially over 27\arcsec\  to increase the signal-to-noise ratio.
After reduction, the wavelength range was 4800 - 5750 \AA\   in the blue
arm spectrum and 6070 - 7020 \AA\   in the red arm, with a scale of
0.55 \AA\  pixel$^{-1}$.  Because no spectrophotometric standard was
observed, only line ratios of closely spaced lines are meaningful.

\section{Results}

\subsection{\ion{H}{1} Distribution}

The ATCA cubes do
not show emission at 1275 km s$^{-1}$, a peak which was originally
detected in the Parkes observations (see Fig.~\ref{fig:f1}).  In the naturally weighted cube (Fig.~\ref{fig:f2}), the first
emission appears at approximately 1635 km s$^{-1}$ and continues to
1675 km s$^{-1}$, covering only 40 km s$^{-1}$.   
There is no emission detected between 1675 and 1755 km s$^{-1}$, and
then emission appears again in the same general region as the first
source.  This emission is stronger and extends from 1760 km s$^{-1}$ to
1840 km s$^{-1}$.   It has a regular structure and is relatively
compact.  The optical velocity of FCC 35 is about 1814 \kms\ (see \S
3.4) so we identify this peak with FCC 35 itself (this source is also
coincident with the optical position of FCC 35 (see Fig.~\ref{fig:f3})), while classifying the first
source as an \ion{H}{1} cloud, separated from FCC 35 by 140 km s$^{-1}$ (peak to peak). 
The \ion{H}{1} cloud has an irregular ring-like structure (see Fig.~\ref{fig:f2}) and no
optical counterpart (see \S3.3). 
 In the uniformly weighted cube, the overall structure of FCC 35 does
not change, but the \ion{H}{1} cloud is resolved out.

 The ATCA fluxes were
obtained using the task BLANK in AIPS to select the channels and
regions which contained emission.  The flux densities for these regions
were then summed together and multiplied by the channel spacing.  For
FCC 35, the ATCA flux is approximately equal to the Parkes' value (see Table~\ref{tab:t3}),
resulting in an \ion{H}{1} mass of 2.2 
$\times$ $10^8$ $M_{\odot}$ and a maximum \ion{H}{1} column density of 1.1 $\times$ $10^{20}$ cm$^{-2}$.  
The ATCA observations of the extended \ion{H}{1} cloud are most likely missing flux due to the absence of
short spacings.  The single-dish
observations give an integrated \ion{H}{1} flux of 2.9 Jy km s$^{-1}$ for the
\ion{H}{1} cloud, while the ATCA gives a value of 1.4 Jy km s$^{-1}$
(see Fig.~\ref{fig:f1}; values are listed in Table~\ref{tab:t4}). 
We adopted the Parkes flux as the most accurate value, giving an \ion{H}{1}
mass of 2.2 $\times$ $10^8$ $M_{\odot}$, or equal to FCC 35.  The maximum \ion{H}{1} 
column density for the \ion{H}{1} cloud using the ATCA data is 2.0 $\times$ $10^{19}$
cm$^{-2}$. 
The 1275 \kms\ peak may be an extended
\ion{H}{1} source which was lost in the ATCA data due to missing {\em uv} spacings.  Future work should 
include observations with more compact configurations to attempt to recover this source.

Moment maps were made of both the \ion{H}{1} cloud and FCC 35 using the MOMNT
task within AIPS and the naturally weighted data cube.  The channels
ranging from 1761 - 1847 km s$^{-1}$ were used to create the moment maps of
FCC 35, and the channels ranging from 1629 - 1675 km s$^{-1}$ were used for the \ion{H}{1} cloud.   
Data greater than and equal to 3 mJy/beam ($\approx 3\sigma$) were included in the maps.  In
Figs.~\ref{fig:f3} and \ref{fig:f4} we show the \ion{H}{1} integrated intensity
contour maps of FCC 35 and the \ion{H}{1} cloud overlaid on an optical
image.  From these pictures it is apparent that the high velocity
source is associated with FCC 35, while the low velocity source does not
appear to have an optical counterpart.  The extended nature of the \ion{H}{1}
is also evident in this figure. 

\subsection{\ion{H}{1} Kinematics}
 The velocity field of FCC 35 is fairly regular, as shown in Fig.~\ref{fig:f5}.  
The \ion{H}{1} cloud does not have any obvious kinematical structure (Fig.~\ref{fig:f6}).  An \ion{H}{1} rotation curve 
was derived for FCC 35 using the tilted ring algorithm ROCUR within AIPS
(Begeman 1987) and a first moment map created with the uniformly weighted cube (for
increased spatial resolution).  
 The galaxy was modeled as a 
succession of independent rings, each defined by its central coordinates, 
systemic velocity, inclination, position angle and rotational velocity.  
The number of rings used in the 
analysis was limited by the large size of the beam relative to FCC 35's
angular diameter.  We oversampled the curve to include several additional points
and kept the central coordinates fixed at the optical center of the
galaxy.  The systemic velocity was fixed at 1800 km s$^{-1}$ after the first run, and the inclination,
position angle and rotational velocity were varied according to ROCUR's analysis.  After
several trial runs, the inclination and position angle were fixed at 13\arcdeg~and 62\arcdeg~respectively, 
and a rotational velocity was obtained for
each ring (Fig.~\ref{fig:f7}).  The entire galaxy was used in this analysis.
Rotation curves were
also obtained for the receding and approaching halves of the galaxy separately, but
they differed only slightly from Fig.~\ref{fig:f7}.

Fig.~\ref{fig:f7} suggests that the rotation curve may be
turning over at our last data point, with a maximum rotational velocity
of $\approx$ 21 km s$^{-1}$ at a radius of 54\arcsec.  Assuming
circular motion, we derive an enclosed mass of $4.8 \times 10^{8}$
M$_{\odot}$ at this radius.  This is probably an underestimate, because
pressure support is usually significant in BCDs (e.g. Meurer et al. 1996; Hunter et al. 1996, 1994).  Indeed, for FCC 35 the ratio of
V$_{rot}^{2}$ to $\sigma_{v}^{2}$ in the outer disk indicates that random
motions contribute greater than 20\% of the support and a cursory examination 
with the AIPS tasks XGAUS and MOMNT suggest an increase in 
linewidth towards the center of the galaxy.
In light of this, we attempted to account for pressure support by calculating an asymmetric
drift correction (Fig.~\ref{fig:f8}).  

The asymmetric drift correction was determined with the method described by Meurer et
al.~(1996), using the radial profile of the \ion{H}{1} surface density and
velocity dispersion.   Our beam size limits the analysis, but adding the 
 correction to our original
rotation curve gives a (corrected) circular velocity curve (Fig.~\ref{fig:f8}; Table~\ref{tab:t5}) which peaks at 
$\approx$ 40\arcsec\ (30 km s$^{-1}$) and subsequently shows a Keplerian decline. 
This suggests that the mass distribution of this galaxy may be
severely truncated. 
Fig.~\ref{fig:f8} shows that the corrected velocity at 60\arcsec\ is the same as our 
uncorrected rotational velocity at this radius, so our mass estimate is 
unaffected.  However, it now represents the total mass of FCC 35 (rather than the
mass enclosed within our last measured point).  Adopting 4.8 $\times$ 10$^{8}$ M$_\odot$ as the total mass 
gives an M$_{tot}$/L$_{\rm V}$ ratio of approximately 3 M$_{\odot}$/L$_{\odot}$.  
  We have not attempted a detailed mass model at this stage due to the large beamsize.  
We defer this analysis until higher resolution
\ion{H}{1} observations are obtained.  However, the results at this stage seem clear:
a Keplerian decline in the velocity curve, which indicates a truncation of mass, low
dark matter content and a low M$_{tot}$/L$_{\rm V}$.

\subsection{Optical Photometry} 
FCC 35's U (Fig.~\ref{fig:f9}a) and V band images
 reveal an off-center nucleus which is not 
present in the I band image (Fig.~\ref{fig:f9}b).  There is no optical 
counterpart for the HI cloud (Fig.~\ref{fig:f4}) (any galaxy with a
surface brightness of V = 26 mag/arcsec$^2$ or M$_V$ = -14 would have
been detected).  The photometric 
calibration for these images was derived 
using Graham (1982) and Landolt (1992) standards (see also 
Bessell 1995) and DAOPHOT in IRAF.   Elliptical 
isophotes were fit to each image using GASP and the remaining data processing was completed within 
the SFOTO package (Han 1991).  The derived magnitudes were corrected for atmospheric and 
galactic extinction.  Elliptical apertures give a
$(U-V)_{22\farcs5}$ = 0.1 and $(U-V)_{14\arcsec}$ = 0.0.  
The total V band magnitude quoted in Table~\ref{tab:t3} was obtained by assuming an exponential
light profile in the outer parts of the disk.  The U band
surface brightness profile can be found in Fig.~\ref{fig:f10}, with the
V and I band profiles behaving similarly.  All of the profiles are tabulated in
Table~\ref{tab:t6}.  Fig.~\ref{fig:f10} appears to
flatten out at the center of the galaxy, but it may be
somewhat ill-determined here due to the off-center nucleus
(see Fig.~\ref{fig:f9}a).  A separate analysis was carried out on the 
galaxy's nucleus using circular apertures with the task PHOT in IRAF.  
This analysis reveals successively bluer colors with decreasing radius 
(Table~\ref{tab:t7}).  As shown in Fig.~\ref{fig:f11} (Schr\"oder 1995), 
FCC 35 is one of the bluest galaxies in the Fornax 
cluster, reflecting 
its star formation activity.

\subsection{Optical Spectrum}

FCC 35's spectrum (Fig.~\ref{fig:f12}) shows strong H${\alpha}$, 
[OIII] $\lambda\lambda 4959, 5007$ and H${\beta}$ emission. The [NII] $\lambda 
6584$ line is marginally detected and [SII] $\lambda\lambda 6716, 
6731$ and HeI $\lambda 6678$ are weak but present.  The ratio
[NII]/H${\alpha}$ has an upper limit of 0.04 and [OIII]/H${\beta}$ has 
a value of approximately 4.8.  Using theoretical photoionization models of
extragalactic \ion{H}{2} regions (Dopita \& Evans 1986; Sutherland \& Dopita
1993), we derive an upper limit on the metallicity of 0.25 Z$_{\odot}$,
which is normal for the absolute magnitude of FCC 35 (see Table~\ref{tab:t3}).  This
value is typical for BCDs (e.g. II Zw 40; Walsh \& Roy 1993), but nowhere near as low as I Zw 18 (Masegosa, Moles
\& Campos-Aguilar 1994).
The optical emission lines are extremely narrow. The optical heliocentric
velocity is $\approx$ 1814 km s$^{-1}$, corresponding to the higher 
velocity \ion{H}{1} peak.

\section{Discussion}
\subsection{The \ion{H}{1} Cloud}

A position-velocity cut through the center of both FCC 35 and the \ion{H}{1}
cloud (Fig.~\ref{fig:f13}) shows that despite the spatial overlap
, the two components are not connected in
velocity space.  
The cloud may either be a unique source 
within the Fornax cluster or a foreground or background \ion{H}{1}
object.  The former seems to be the most likely considering the velocity
of the cloud (V$_{r,\odot}$ = 1658 km s$^{-1}$) and the mean heliocentric
velocity of the Fornax cluster ($\langle{v}\rangle = 1450 \pm$ 34 km s$^{-1}$, $\sigma_v$ = 350 km s$^{-1}$; Held \&
Mould 1994). 

The Fornax cluster has a central number density of 500 galaxies Mpc$^{-3}$
(Ferguson 1989), and tidal debris from interactions should be expected
(e.g. Theuns \& Warren 1997).  Some of the intergalactic material may
be primordial, but in a dense cluster it is likely to have arisen from 
galactic harassment (Moore et al. 1996).  
This is especially true when considering the proximity of our \ion{H}{1} cloud to the center of the
Fornax cluster ($03^{h}35^{m}, -35.7^{\circ}$; Ferguson 1989).  
The interaction which formed the cloud may not have involved the galaxy FCC 35, and the
cloud may simply be ``passing
by'' at this stage.  Its structure could be affected by FCC 35's presence
(see Figs.~\ref{fig:f2} \&~\ref{fig:f4}), but this is difficult to
confirm due to the irregular kinematics of the cloud
and the sensitivity of its inferred structure on the weighting used in
the reduction.

It is also possible that the \ion{H}{1} cloud is a remnant of an interaction
in which FCC 35 was involved.  We note that NGC 1316C is located only
6$^{\prime}$ away (in projection) from FCC 35 ($\Delta$V = 150 km s$^{-1}$) and the two
could have interacted in the past.  It is conceivable
that FCC 35 and the cloud were a single object which was ripped apart 
tidally into two parts of comparable mass.  This seems unlikely, however, 
considering the regular and compact structure of FCC 35.

Yet another possibility is that the \ion{H}{1} cloud {\em and} FCC 35 formed
through an interaction between two spirals.  Dwarf galaxies and massive
\ion{H}{1} clouds have been predicted to form (Barnes \& Hernquist 1992;
Elmegreen et al. 1993) and observed forming (Schweizer 1978;
Mirabel et al 1991) in the tidal tails which arise from these interactions.
The \ion{H}{1} cloud could then be the remaining gas from a tidal tail.  This
possibility will be discussed further in the next section.

\subsection{FCC 35}

The amount of neutral hydrogen in FCC 35 ($M_{HI}/M_{Tot}$ = 0.5), together with 
the optical data, presents a picture of a blue compact dwarf (BCD) or \ion{H}{2} galaxy.  
Exponential surface brightness profiles are typical of BCDs 
(see Fig.~\ref{fig:f10}), as are offset nuclei (Fig.~\ref{fig:f9}a; Drinkwater \& Hardy 1991).
The spectrum of FCC 35 is also similar to that of the general BCD population, with
relatively low metallicity and strong narrow emission lines (Fig.~\ref{fig:f12}; Izotov et al. 1997; Masegisa et al. 1994; Walsh \& Roy 1993; Thuan \& Martin 1981).  
The strong H${\alpha}$ and [OIII] emission lines are a signature of the star formation 
activity in FCC 35, as are the blue colors towards the galaxy's nucleus 
(see Table~\ref{tab:t7}).

The formation of BCDs is still not understood. Interaction may be
responsible for a significant fraction, but there are also explanations
based upon an evolutionary sequence among dwarf galaxies.   Davies \&
Phillipps (1988) propose a sequence,
dI$\leftrightarrow$BCD$\leftrightarrow$dE, which involves
repeatedly induced star formation bursts and explains the similarities between
different types of dwarf galaxies.  The trigger of the BCD phenomenon is described
by Gordon \& Gottesman (1981).  They find the majority of dwarf
irregulars to have an extended \ion{H}{1} halo and suggest that the infall of
this halo fuels the star formation bursts.  FCC 35 does have an extended
\ion{H}{1} halo, but the presence of the \ion{H}{1} cloud suggests that this star 
formation burst is interaction related.

One type of interaction which has been found to produce BCDs and
intergalactic \ion{H}{1} clouds is the interaction between two spiral
galaxies.  The tidal tails formed as a result of these interactions can
extend for hundreds of kpc (Hibbard \& Van Gorkom 1996)
and create objects of up to 10$^9$ M$_{\odot}$
(Elmegreen et al. 1993).  The tidal features often
have low mass-to-light ratios (Hibbard \& Van Gorkom 1996),
and the models of Barnes and Hernquist (1992) 
predict that these objects would have very little dark matter.
Elmegreen et al. (1993) also predict that dwarf galaxies
formed as a result of this type of interaction should contain old stars from
the original disks plus new stars from the interaction-induced
star formation bursts. FCC 35's upper limit on the ionized gas abundance (Z $\leq$ 0.25 
Z$_{\odot}$) is consistent with tidal formation from the outer 
disk of a spiral galaxy.  Indeed, FCC 35 fulfills many of the criteria
related to the spiral-spiral interaction scenario
(see Table~\ref{tab:t3}). It has a relatively low $M_{tot}$/$L_V$ and a (corrected)
rotation curve (Fig.~\ref{fig:f8}) which indicates a truncated mass distribution and (presumably) small amounts of dark matter.  
However, if this is the origin of FCC 35, we would
perhaps expect to observe more remnant \ion{H}{1} in the surrounding region. This 
was not apparent in the channel maps obtained within the 43$^{\prime}$
primary beam of the ATCA.

FCC 35's star formation burst could also have been induced through an
interaction with its closest neighbor, NGC 1316C, which has not been detected
in \ion{H}{1}.  However, this scenario appears unlikely when the age of
FCC 35's star burst is taken into account.  Its color,  (U-V)$_{26\arcsec}$ = 0.1, corresponds
to an age of about 10$^{7}$ years (Larson \& Tinsley 1978).  If FCC 35's star formation burst had been
directly triggered by an interaction with NGC 1316C, the relative speed
of NGC 1316C would need to be at least 3000 km s$^{-1}$. This is
excessive given the low velocity dispersion of the Fornax cluster.  However, we
recall that internal motions resulting from interactions can
induce later star formation bursts (Noguchi 1991), so our arguments do not
conclusively exclude this possibility or the spiral-spiral interaction
scenario.

The most plausible interaction-related cause for the star formation burst in FCC 35
is that the \ion{H}{1} cloud, whatever its origin, has triggered it.  
Its relative mass and its proximity in space and velocity
make it likely that the cloud is interacting with FCC 35.  This situation is
not uncommon: Taylor et al. (1994) find that galaxies with \ion{H}{1}
companions tend to have a very low mass-to-light ratios and the mass of
the companion is often only an order of magnitude smaller than the mass
of the galaxy (see also Walter et al. 1997). The relative velocity, projected separation, and masses
of the \ion{H}{1} cloud and FCC 35 clearly show that they are unbound, so
it seems likely that they will drift apart as FCC 35 fades into a low surface brightness 
or irregular dwarf galaxy.

\section{Summary}
We have obtained radio synthesis data, complemented by optical imaging
and spectroscopy, of FCC 35, a Fornax member galaxy which has an unusual single-dish \ion{H}{1} profile.
We discovered a rotationally supported \ion{H}{1} disk associated with FCC 35 and
an irregularly shaped \ion{H}{1} cloud.  FCC 35 and the \ion{H}{1} cloud have
approximately equal \ion{H}{1} masses and the two sources overlap spatially
yet are completely separate in velocity.  The cloud has no kinematical
structure, while FCC 35 has a rotation curve which indicates
a truncated mass distribution.  U and V band images depict an offset
nucleus in FCC 35 and no optical counterpart for the \ion{H}{1} cloud.  
FCC 35 has very blue $U-V$ colors and
strong H$_{\alpha}$ and [OIII] emission lines, indicating it is a source of active
star formation. 
The data indicate that the \ion{H}{1} cloud is most likely triggering the star formation burst
in FCC 35. 
Whatever FCC 35's origin and fate, it is an interesting source of
information on the evolution of dwarf galaxies in a cluster
environment.  Further work should investigate FCC 35's surrounding
environment and explore its possible influence on the
gas which has become FCC 35.  Deep \ion{H}{1} searches for intergalactic clouds in 
 clusters should be 
persued in Fornax as well as other clusters. 

\acknowledgements
MEP and MB acknowledge the support of an Australian DEETYA 
Overseas Postgraduate Research Scholarship.  MB also acknowledges the support of a Canadian
NSRC Postgraduate Scholarship.
We thank the staff of Mount Stromlo and Siding Spring Observatories and of the
Australia Telescope National Facility for their assistance during and after
the observations.  In particular, Richard Gooch for his help in maintaining the
Karma Visualization Software, Helmut Jerjen for his help with the observations and Mike Dopita
for the use of the MAPPINGS program.
We also thank the anonymous referee for useful comments.
The Australia Telescope is funded by the Commonwealth of Australia for operation 
as a National Facility managed by CSIRO. 
The Digitized Sky Surveys were produced at the Space Telescope Science
Institute under U.S. Government grant NAG W-2166. The images of these surveys
are based on photographic data obtained using the Oschin Schmidt Telescope on
Palomar Mountain and the UK Schmidt Telescope. The plates were processed into
the present compressed digital form with the permission of these institutions.

\clearpage

\begin{figure}
\caption{FCC 35's 21 cm line profile as measured with the Parkes Telescope (bold line) 
and the ATCA (thin line).  The Parkes spectrum has been corrected
for the offset used in the observations (a correction factor of 1.38).}
\label{fig:f1}
\end{figure}

\begin{figure}
\caption{Naturally weighted channel maps for the 21 cm ATCA data.  The heliocentric velocity 
is shown in the upper left corner of each
channel and the cross marks the optical center of FCC 35.  The beam is shown in the lower left corner of the
first channel.
Contours range from 5 - 40 mJy/beam in steps of 6 mJy/beam, with an rms per channel of 1.2 mJy/beam.}
\label{fig:f2} 
\end{figure}

\begin{figure}
\caption{DSS image overlaid on the \ion{H}{1} integrated intensity contours of FCC 35.  
The image is 10$^\prime \times 10^\prime$ and the contours range from 3 - 170 mJy/beam in
steps of 16 mJy/beam. 
The beam is shown in the lower left corner.}
\label{fig:f3}
\end{figure}

\begin{figure}
\caption{DSS image overlaid on the \ion{H}{1} integrated intensity contours of the \ion{H}{1} 
cloud.  The image is 10$^\prime \times 10^\prime$ and the contours range from 3 -
34.5 mJy/beam in steps of 4.5 mJy/beam.   
The beam is shown in the lower left corner.}
\label{fig:f4}
\end{figure}

\begin{figure}
\caption{Velocity contours of FCC 35.  Several of the contours are labeled in km s$^{-1}$.}
\label{fig:f5}
\end{figure}

\begin{figure}
\caption{Velocity contours of the \ion{H}{1} cloud.  The contours
 have no obvious gradient and are labeled in km s$^{-1}$.}
\label{fig:f6}
\end{figure}

\begin{figure}
\caption{Uncorrected rotation curve of FCC 35 using both sides of the galaxy.}
\label{fig:f7}
\end{figure}

\begin{figure}
\caption{Corrected rotation curve of FCC 35 (solid circles).  The open circles show the \ion{H}{1} rotation
curve from Fig.~\ref{fig:f7}, 
and the solid line represents the assymetric drift correction.
 The data are tabulated in
Table~\ref{tab:t5}.}
\label{fig:f8}
\end{figure}

\begin{figure}
\caption{a) U band, and b) I band images of FCC 35.  Each image is 40\arcsec $\times$ 45\arcsec.}
\label{fig:f9}
\end{figure}

\begin{figure}
\caption{U band surface brightness profile of FCC 35.  The data for the U, V and I band surface
brightness profiles are tabulated in Table~\ref{tab:t6}.}
\label{fig:f10}
\end{figure}

\begin{figure}
\caption{Distribution of integrated galaxy colors within the Fornax cluster.  FCC 35 is 
marked as a solid circle. (See Schr\"oder 1995 for more information on the photometry.)}
\label{fig:f11}
\end{figure}

\begin{figure}
\caption{a) Blue spectrum of FCC 35, showing $H_{\beta}$ and [OIII] $\lambda\lambda$ 
4959, 5007. b) Red spectrum of FCC 35 showing $H_{\alpha}$, [NII] $\lambda$6584, 
HeI $\lambda$6678, and [SII] $\lambda\lambda$6716, 6731.  No other lines are seen in the spectra.}
\label{fig:f12}
\end{figure}

\begin{figure}
\caption{Position-velocity slice through the centers of both FCC 35 and 
the \ion{H}{1} cloud (PA=91$^\circ$, naturally weighted cube).}
\label{fig:f13}
\end{figure}

\clearpage

\begin{table}
\dummytable\label{tab:t1}
\end{table}

\begin{table}
\dummytable\label{tab:t2}
\end{table}

\begin{table}
\dummytable\label{tab:t3}
\end{table}

\begin{table}
\dummytable\label{tab:t4}
\end{table}

\begin{table}
\dummytable\label{tab:t5}
\end{table}

\begin{table}
\dummytable\label{tab:t6}
\end{table}

\begin{table}
\dummytable\label{tab:t7}
\end{table}

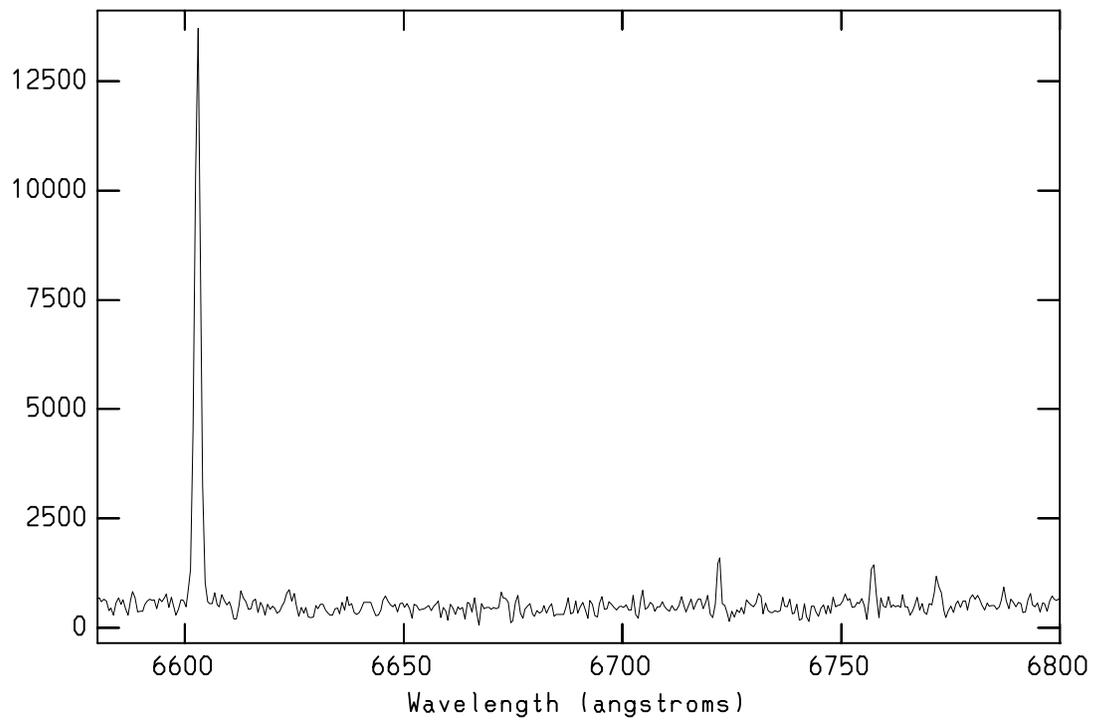
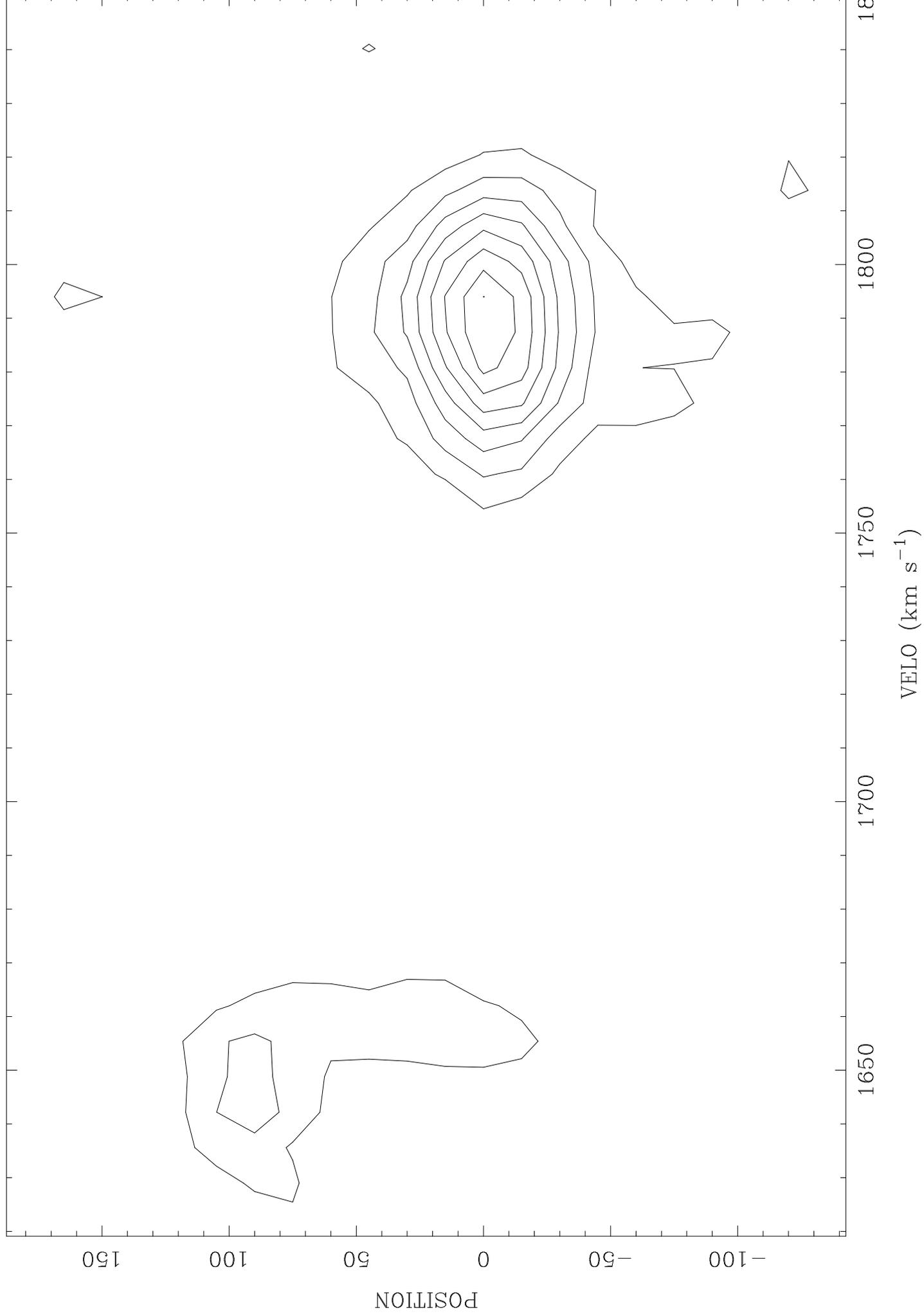
\end{document}